\documentclass{elsart}

\usepackage{epsfig}
\usepackage{amssymb}

\begin{document}

\begin{frontmatter}

\title{Shot noise as a tool to probe an electron energy distribution}
\author{O. M. Bulashenko\corauthref{cor1}},
\corauth[cor1]{Corresponding author.}
\ead{oleg@ffn.ub.es}
\author{J. M. Rub\'{\i}}
\address{Departament de F\'{\i}sica Fonamental, Universitat de Barcelona,
Av.\ Diagonal 647, E-08028 Barcelona, Spain}

\begin{abstract}
We discuss the possibility to employ the shot-noise measurements 
for the analysis of the energy resolved ballistic currents.
Coulomb interactions play an essential role in this technique,
since they lead to the shot-noise-suppression level which depends on 
the details of the energy profile.
\end{abstract}

\end{frontmatter}

\section{Introduction}

Recently, shot-noise measurements are emerging as an important tool to obtain 
information on the transport properties and interactions among carriers
in quantum structures \cite{landauer98,blanter00}.
Since Coulomb repulsion between electrons and their fermionic nature can 
regulate their motion, this effect may be detected in the shot-noise 
reduction (in respect to the Poissonian value), but cannot be deduced from 
time-averaged dc measurements.
In particular, shot noise has been used as a tool to probe: 
fractional charge \cite{frac}, effective superconducting charge \cite{superc}, 
quantum transmission modes in atomic-size contacts \cite{atomcont}, 
mechanisms of tunneling \cite{tun}, etc. (see, e.g., recent review 
\cite{blanter00}).

In this contribution, we discuss the possibility to employ the shot-noise
measurements to obtain information on the energy distribution of 
nonequilibrium carriers injected from a contact emitter.
Using ballistic electrons to study nanoscale structures has recently been 
a very active research area.
In a usual technique called ``hot-electron spectroscopy'' 
\cite{heiblum85,xie99}, 
carriers injected from an emitter contact are analyzed in a collector 
contact by means of a barrier which is transparent only for the carriers
having energy greater than the barrier height. 
By changing the bias on the collector barrier, the electron energy profile 
can be analyzed. This technique requires the design of a special collector 
filter for getting information on the electron energies.
Here, we propose a simpler method which does not require a design of 
the filter, rather it employs a ``natural'' filter: the potential barrier 
that appears due to an injected space charge.
This space charge limits the current producing the resistance effect by means 
of a barrier, which reflects a part of the injected carriers back 
to the emitter.
The height of the barrier depends on the screening parameter of the material, 
and it varies with the external bias.
The essential difference with the case of a fixed barrier is that 
the space-charge barrier fluctuates in time and produces the long-range 
Coulomb correlations between the transmitted electrons,
that leads to the significant suppression of shot noise registered 
at the collector contact. 
The level of suppression depends drastically on the energy profile 
of the injected carriers \cite{prb00b}, while the time-averaged quantities 
(the mean current, conductance, etc.) do not.

\begin{figure}[b]
\epsfxsize=12.0cm
\centerline{
\epsfbox{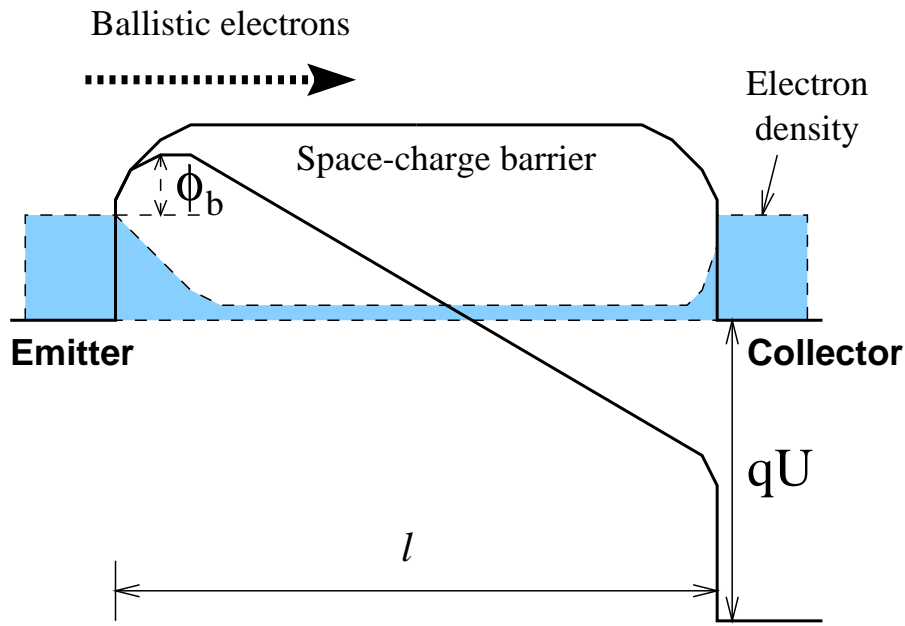}
}
\protect\vspace{1cm}
\caption
{\footnotesize
Band-energy diagram for a ballistic conductor under a space-charge-limited
regime. At equilibrium, the potential space-charge barrier is symmetrical,
while under the applied bias $U$ it is closer to the emitter contact,
and its magnitude $\Phi_b$ decreases with bias. 
The electron density (shadowed regions) increases near the emitter contact
when the bias is enhanced.
}
\end{figure}

\section{Shot-noise suppression by Coulomb correlations}

We consider the current injection into a ballistic conductor at high biases
(Fig.\ 1).
To adequately describe the transport and noise under the nonlinear 
far-from-equilibrium conditions, one has to solve the transport equation 
coupled self-consistently with a Poisson equation \cite{prb00b}. 
The self-consistent built-in field determines the potential barrier $\Phi_b$, 
at which electrons are either reflected or transmitted depending on 
their energy.
The steady-state ballistic current $I$ is determined by the electrons 
passed over the barrier (tunneling is neglected):

\begin{equation} \label{I}
I = \frac{q}{2\pi\hbar} 
\int_{\Phi_b}^{\infty} N(\varepsilon)\ d\varepsilon,
\end{equation}
where $q$ is the electron charge, 
$N(\varepsilon)= A
\int f(\varepsilon,{\bf k}_{\perp}) d {\bf k}_{\perp}/(2\pi)^{d-1}$
is the number of occupied transversal modes,
$A$ is the cross-sectional area, $d$ is the dimension of a momentum space and 
$f(\varepsilon,{\bf k}_{\perp})$ 
is the occupation number of a quantum state of a longitudinal energy 
$\varepsilon$ and a transverse momentum ${\bf k}_{\perp}$ at the emitter.
Note that the differential conductance for this kind of emitter is given by 
$G = c\ (q^2/2\pi\hbar) N(\Phi_b)$,
where $c(U,\ell)$ is the Coulomb interaction factor dependent on the bias $U$
and the length of the ballistic region $\ell$.

Since there is no scattering, the occupation of transversal modes is conserved
during the motion, and the only source of noise is the fluctuations of
the occupation numbers at the emitter modulated by the potential-barrier
fluctuations $\delta\Phi_b$. The barrier fluctuations induce the long-range 
Coulomb correlations, and they are found from the Poisson equation 
\cite{prb00b}.
The result may be summarized by introducing the current-fluctuation transfer 
function $\gamma(\varepsilon)$ as follows: 

\begin{equation} \label{dI}
\delta I = \frac{q}{2\pi\hbar} \left[ 
\int_{\Phi_b}^{\infty} \delta N(\varepsilon)\ d\varepsilon
- N(\Phi_b)\ \delta \Phi_b  \right] \quad
= \quad \frac{q}{2\pi\hbar} \int_{\Phi_b}^{\infty} \gamma(\varepsilon)\ 
\delta N(\varepsilon)\ d\varepsilon.
\end{equation}
As a result of the long-range Coulomb interactions, the transmission
for different energies differs by the factor $\gamma(\varepsilon)$.
Thus, the low-frequency current-noise spectral density is 

\begin{eqnarray} \label{SI}
S_I = \int_{\Phi_b}^{\infty} \gamma^2(\varepsilon)\
K(\varepsilon)\ d\varepsilon,
\end{eqnarray}
where $K$ is given through the correlator
$\langle \delta N(\varepsilon)\delta N(\varepsilon') \rangle
=(q/2\pi\hbar)^2 K(\varepsilon)\ (\Delta f)\ \delta(\varepsilon-\varepsilon')$,
with $\Delta f$ the frequency bandwidth. At high biases $qU\gg \Phi_b$, 
we find the asymptotic formula for the function $\gamma$

\begin{eqnarray} \label{gamma}
\gamma(\varepsilon) = \frac{3}{\sqrt{qU}} 
\left[ \sqrt{\varepsilon-\Phi_b} - v_{\Delta} \right], 
\end{eqnarray}
where

\begin{eqnarray} \label{vd}
v_{\Delta} = \frac{1}{N(\Phi_b)} \int_{\Phi_b}^{\infty} 
\left(- \frac{\partial N}{\partial \varepsilon}
\right) \sqrt{\varepsilon-\Phi_b}\ d\varepsilon.
\end{eqnarray}
It is important to highlight that the current noise (\ref{SI}) depends on
the details of the distribution function $N(\varepsilon)$ at the emitter 
through the parameter $v_{\Delta}$. The function $\gamma(\varepsilon)$ plays
the role of the energy-resolved shot-noise suppression factor.

\section{Example}

To illustrate the implementation of the results, we consider the emitter
in which the injection energy profile has a peak at energy $\varepsilon_0$
superimposed on a wide background of 3D Fermi-Dirac (FD) electrons 
(see Inset in Fig.~2). Under the assumption that the width of the peak 
is narrow on the scale of the temperature $T$, the number of the occupied 
modes for each longitudinal energy $\varepsilon$ can be written as

\begin{equation} \label{nFD}
N(\varepsilon)= \frac{m k_B T}{2\pi\hbar^2} \Bigl[
\ln \{1+\exp[(\varepsilon_F -\varepsilon)/k_BT]\} + \tilde{a}\ k_BT\ 
\delta(\varepsilon-\varepsilon_0) \Bigr]
\end{equation}
where $\varepsilon_F$ is the Fermi energy, 
and $\tilde{a}$ is the dimensionless peak magnitude. 
It is known that the mean current for ballistic injection 
in the limit of high biases is described by the Child law independently 
of the distribution function \cite{prb00b}. 
For the distribution (\ref{nFD}), the steady-state current (\ref{I}) 
is obtained as ($\varepsilon_0>\Phi_b$ is assumed)

\begin{equation} \label{IFD}
I = \frac{q}{2\pi\hbar} N_S \frac{k_BT}{\xi} 
[F_1(\xi-\Phi_b/k_BT) + \tilde{a}],
\end{equation}
where $N_S=k_F^2A/(4\pi)$, $k_F^2=2m\varepsilon_F/\hbar^2$, 
$\xi=\varepsilon_F/k_BT$, and we denote the Fermi-Dirac integrals of index $j$
by 
$F_j(\alpha)=\int_0^{\infty}x^j [1+\exp(x-\alpha)]^{-1} dx/\Gamma(j+1)$.
It is seen that the information on the peak position $\varepsilon_0$ is lost
in the time-averaged current $I$. In contrast, the noise is sensitive to both 
the peak position and its magnitude, as will be shown below.

Assuming that the peak electrons are uncorrelated (Poissonian), while
the FD electrons are correlated by the Pauli exclusion, we obtain 
the correlation function for the injected carriers \cite{prb01}

\begin{equation} \label{KF3D}
K(\varepsilon) = \frac{2G_{\rm S}}{\xi}\ [ F_{-1}(\xi-\varepsilon/k_BT)
+ \tilde{a}\ k_BT\ \delta(\varepsilon-\varepsilon_0) ],
\end{equation}
where $G_{\rm S}=(q^2/2\pi\hbar)N_S$ is the Sharvin conductance. 
Thus for the current noise we obtain

\begin{equation} \label{SIFD}
S_I =  2qI_{bg}\ \frac{9k_BT}{qU}\ 
[1 - \sqrt{\pi} w g_1 + w^2 g_2 + a\ (\sqrt{\varepsilon_{\delta}}-w)^2 ],
\end{equation}
where $I_{bg}$=$I(\tilde{a}=0)$ is the background current,
$w\equiv v_{\Delta}/\sqrt{k_BT}=
(\sqrt{\pi}g_1 + a/\sqrt{\varepsilon_{\delta}})/(2g_2)$,
$a$=$\tilde{a}/F_1(\alpha)$ is the ratio between the current of the peak
and the current of the background,
$g_1$=$F_{1/2}(\alpha)/F_1(\alpha)$ and
$g_2$=$F_0(\alpha)/F_1(\alpha)$ are the coefficients
dependent on the degree of degeneracy of injected electrons
($g_1$=$g_2$=1 at high temperatures for nondegenerate electrons), and 
$\alpha=\xi-\Phi_b/k_BT$ and 
$\varepsilon_{\delta}=(\varepsilon_0-\Phi_b)/k_BT$ are the positions of the 
Fermi energy and the $\delta$ peak with respect to the potential barrier.
The current noise (\ref{SIFD}) has a form of a shot noise suppressed by Coulomb
interactions. The suppression is enhanced with the bias $U$ and depends on 
the details of the injected distribution, in particular on the peak parameters
$a$ and $\varepsilon_{\delta}$. 
Note that for noninteracting carriers ($\gamma$=$1$), one gets
$S_I=2qI_{bg}(g_2+a)$, that means the information on the peak position
is lost, in a similar way as for the average current (\ref{IFD}).
For this case the noise is a sum of two shot-noise terms: the first one 
is multiplied by the factor $g_2$ (Fermi suppressed Poissonian shot noise 
\cite{prb00b}),
and the second one is the Poissonian noise from the $\delta$ peak.
Thus, the role of the Coulomb interactions is crucial in providing 
an additional information on the injected carriers.

In the experiment, if one is interested in getting information on the peak
electrons, one can register the noise for two cases: with and without
the $\delta$ peak, and then find the ratio 

\begin{equation} \label{beta}
\beta = \frac{S_I}{S_I^{bg}} = 1 + a\ 
\frac{(\sqrt{\varepsilon_{\delta}}-w)^2+a/(4g_2\varepsilon_{\delta})}
{1-(\pi g_1^2)/(4g_2)}.
\end{equation}
It is seen that the parameter $\beta$ is a nonmonotonic function
of the peak position.
We have studied the behavior of $\beta$ when the peak position 
$\varepsilon_{\delta}$ and the peak current $a$ are varied.
The results are illustrated in Fig.~2.
For a fixed $a$, each curve displays a minimum at some energy correspondent
approximately to the energy $\varepsilon_0^*=\Phi_b+v_{\Delta}^2$, 
where the current-fluctuation transfer function vanishes: 
$\gamma(\varepsilon_0^*)$=0.
Since the barrier height $\Phi_b$ is a function of the applied bias,
one can tune by $U$ the $\delta$ peak position to $\varepsilon_0^*$ 
where the noise is minimal, thereby revealing the peak position.
In the opposite case, when the peak position is given, one may get
information on the screening parameter at the space-charge barrier.

\begin{figure}[t]
\epsfxsize=12.0cm
\centerline{
\epsfbox{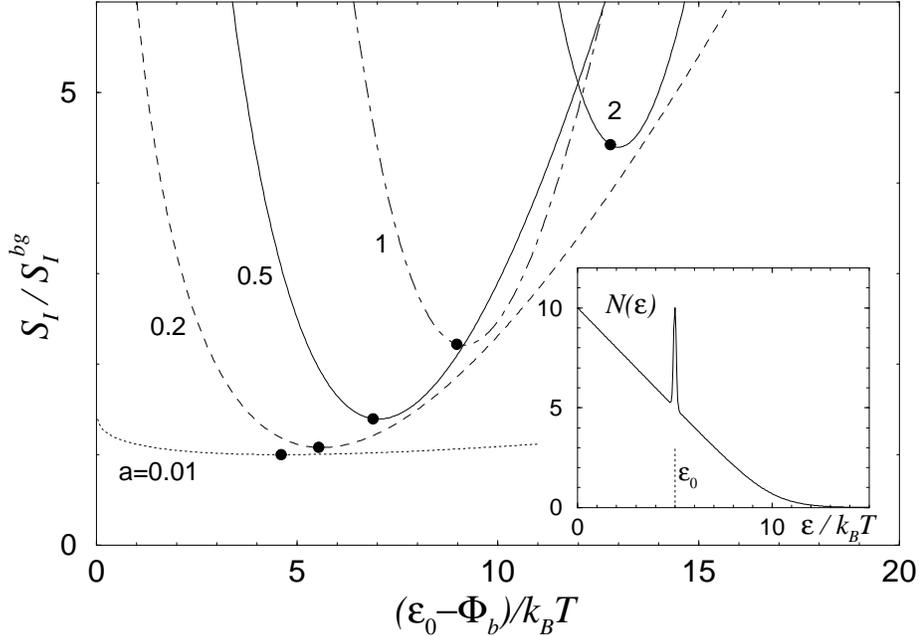}
}
\protect\vspace{1cm}
\caption
{\footnotesize
Current noise produced by Fermi-Dirac electrons with an 
additional $\delta$ peak at $\varepsilon$=$\varepsilon_0$ 
(the distribution is shown in the inset)
with respect to the case when no peak is present.
The ratio $S_I/S_I^{bg}$ is shown as a function of the peak position
$(\varepsilon_0-\Phi_b)/(k_BT)$ for $\alpha$=10 and different magnitudes 
of the peak current $a$.
For each curve, the point near the minimum indicate the energy at which
the energy-resolved shot-noise suppression factor vanishes $\gamma$=0.
}
\end{figure}

Summarizing, we have shown that Coulomb interactions in ballistic structures
are of interest from several points of view:
On the one hand, they lead to the shot-noise suppression 
that may be important for applications.
On the other hand, they give the possibility to use the shot-noise 
measurements as a tool to deduce 
an important information on the properties of nonequilibrium carriers 
in nanoscale structures with hot-electron emitters, 
resonant-tunneling-diode emitters, superlattice emitters, etc., 
not otherwise available from dc measurements.

\end{document}